\documentclass[sigconf]{acmart}

\copyrightyear{2022}
\acmYear{2022}
\setcopyright{acmcopyright}
\acmConference[SIGIR '22]{Proceedings of the 45th International ACM SIGIR Conference on Research and Development in Information Retrieval}{July 11--15, 2022}{Madrid, Spain}
\acmBooktitle{Proceedings of the 45th International ACM SIGIR Conference on Research and Development in Information Retrieval (SIGIR '22), July 11--15, 2022, Madrid, Spain}
\acmPrice{15.00}
\acmDOI{10.1145/3477495.3531772}
\acmISBN{978-1-4503-8732-3/22/07}

\AtBeginDocument{%
  \providecommand\BibTeX{{%
    \normalfont B\kern-0.5em{\scshape i\kern-0.25em b}\kern-0.8em\TeX}}}

\usepackage{bbm}
\usepackage{enumitem}
\usepackage{multirow}
\usepackage{graphicx}
\usepackage{sistyle}
\usepackage[ruled]{algorithm2e}
\usepackage{booktabs}
\usepackage{caption}
\SIthousandsep{,}
\usepackage{makecell}
\usepackage[skip=0pt]{caption}
\usepackage{amsmath}
\usepackage{subfigure}

\newcommand\blfootnote[1]{%
  \begingroup
  \renewcommand\thefootnote{}\footnotetext{#1}%
  \addtocounter{footnote}{-1}%
  \endgroup
}

\begin{document}
\fancyhead{}
\title{Pre-train a Discriminative Text Encoder for Dense Retrieval via Contrastive Span Prediction}


\author{Xinyu Ma, Jiafeng Guo$^{*}$, Ruqing Zhang, Yixing Fan and Xueqi Cheng}
\affiliation{
  \institution{
    CAS Key Lab of Network Data Science and Technology, Institute of Computing Technology, \\ Chinese Academy of Sciences, Beijing, China\\
    University of Chinese Academy of Sciences, Beijing, China\\}
    \country{}
}
\email{{maxinyu17g,guojiafeng,zhangruqing,fanyixing,cxq}@ict.ac.cn}

\renewcommand{\shortauthors}{Xinyu Ma, et al.}

\begin{abstract}
\blfootnote{$^{*}$ Jiafeng Guo is the corresponding author.}

Dense retrieval has shown promising results in many information retrieval (IR) related tasks, whose foundation is high-quality text representation learning for effective search.
Some recent studies have shown that autoencoder-based language models are able to boost the dense retrieval performance using a weak decoder. 
However, we argue that 1) it is not discriminative to decode all the input texts and, 2) even a weak decoder has the bypass effect on the encoder. 
Therefore, in this work, we introduce a novel contrastive span prediction task to pre-train the encoder alone, but still retain the bottleneck ability of the autoencoder.
The key idea is to force the encoder to generate the text representation close to its own random spans while far away from others using a group-wise contrastive loss. 
In this way, we can 1) learn discriminative text representations efficiently with the group-wise contrastive learning over spans and, 2) avoid the bypass effect of the decoder thoroughly. 
Comprehensive experiments over publicly available retrieval benchmark datasets show that our approach can outperform existing pre-training methods for dense retrieval significantly. 
Code and pre-trained models will be available at the URL
\footnote[1]{https://github.com/Albert-Ma/COSTA}.

\end{abstract}

\ccsdesc[500]{Information systems~Information retrieval}

\keywords{Dense Retrieval, Pre-training for IR, Discriminative Representation}


\maketitle

\section{Introduction}\label{sec:intro}

Dense retrieval is receiving increasing interest in recent years from both industrial and academic communities due to its benefits to many IR related tasks, e.g., Web search~\cite{Cai2021SemanticMF,Lin2021PretrainedTF,Fan2021PretrainingMI}, question answering~\cite{Karpukhin2020DensePR,Lee2021LearningDR,zhu2021} and conversational systems~\cite{Gao2022NeuralAT,Yu2021FewShotCD}. 
Without loss of generality, dense retrieval usually utilizes a Siamese or bi-encoder architecture to encode queries and documents into low-dimensional representations to abstract their semantic information \cite{Xiong2021ApproximateNN,Zhan2021OptimizingDR,Khattab2020ColBERTEA,Hofsttter2021EfficientlyTA,Humeau2020PolyencodersAA,Zhan2020RepBERTCT}. 
With the learned representations, a dot-product or cosine function is conducted to measure the similarity between queries and documents. 
In essence, high-quality text representation is the foundation of dense retrieval to support effective search in the representation space.

Taking the pre-trained representation models like BERT \cite{Devlin2019BERT} and RoBERTa \cite{Liu2019RoBERTaAR} as the text encoders have become a popular choice \cite{Xiong2021ApproximateNN,Zhan2021OptimizingDR,Khattab2020ColBERTEA} in dense retrieval.
Beyond these direct applications, there have been some works on the pre-training objectives tailored for dense retrieval \cite{Lee2019LatentRF,Chang2020PretrainingTF}. 
For example, \citet{Chang2020PretrainingTF} presented three pre-training tasks that emphasize different aspects of semantics between queries and documents, including Inverse Cloze Task (ICT), Body First Selection (BFS), and Wiki Link Prediction (WLP). 
As we can see, some tasks even depend on certain special document structures, e.g., hyperlinks.
When applying such pre-trained models to dense retrieval, marginal benefit could be observed on typical benchmark datasets as shown in Section \ref{sec:baseline}.

\begin{figure}
    \centering
    \includegraphics[scale=0.25]{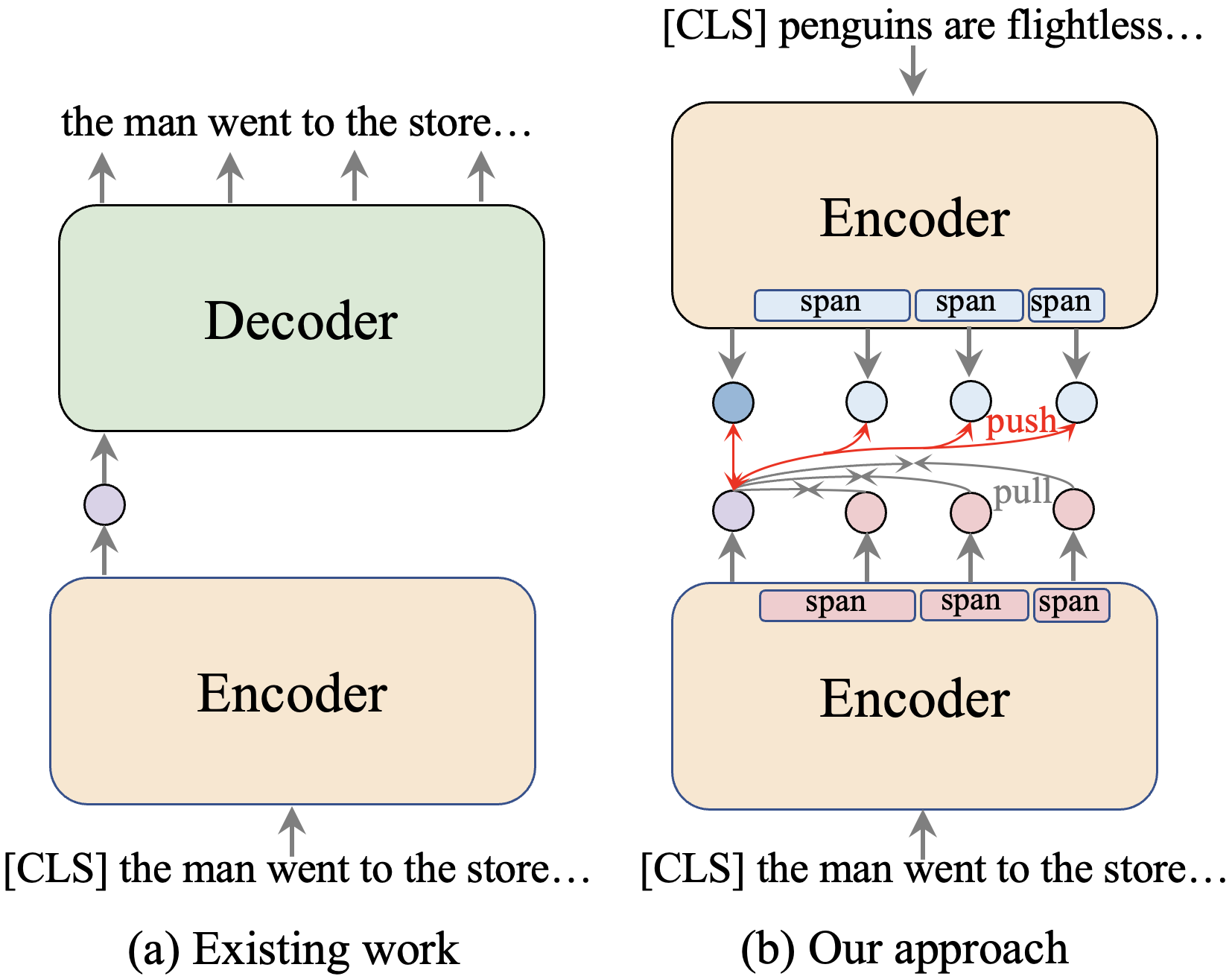}
    \caption{COSTA pre-trains the encoder alone with the contrastive span prediction task while still retaining the bottleneck ability by forcing the encoder to generate the text representation close to its own random spans. Moreover, it enhances the discriminative ability of the encoder by pushing the text representation away from others.}
    \label{fig:ae}
\end{figure}

To boost the dense retrieval performance, recent studies begin to focus on the autoencoder-based language models, which are inspired by the information bottleneck \cite{Tishby2015DeepLA} to force the encoder to provide better text representations \cite{Li2020OptimusOS,lu-etal-2021-seed}. 
As shown in Figure \ref{fig:ae} (a), these methods pair a decoder on top of the encoder and then train the decoder to reconstruct the input texts solely from the representations given by the encoder. 
When generating text in the autoregressive fashion, the model takes not only the encoder's encodings but also the previous tokens as input. 
Through mathematical analysis, \citet{lu-etal-2021-seed} showed that the decoder can bypass the dependency on the encoder via its access to previous predicted tokens, especially when the text is long and the decoder is strong. 
That is, a low reconstruction loss does not mean good text representations.
To address this issue, they proposed to pre-train the autoencoder using a weak decoder, with restricted capacity (i.e., a shallower Transformer) and attention flexibility (i.e., restricting its access to the previous context).

Despite the great success on dense retrieval, the autoencoder-based language approach has several shortcomings: (1) It assumes that each token is equally important to the input text. 
A large proportion of words in the text are common words like the, of, etc, since they are useful for composing the natural language.
Thus, in the auto-regressive decoder, the model tends to decode common words to achieve a low reconstruction loss.
As a result, the discriminative power of representations may be decreased, especially when the text is long with much noisy information.
(2) The decoder, which is useless in the inference stage, is not a necessary part in the training stage either. 
Regardless of how weak the decoder is, it still has access to previous tokens to exploit the natural language patterns. 
Therefore, the bypass effect of the decoder still remains which may largely limit the encoding power of the encoder.

Therefore, in this paper, we propose to drop out the decoder and enforce the information bottleneck by the encoder itself to provide better text representations.
The key idea is to enhance the consistency between semantic representations of the given text and that of its own random spans (i.e., a group) using a group-wise contrastive loss. 
To learn both the bottleneck and the discriminative ability, we introduce the contrastive span prediction task for dense retrieval.
Specifically, as shown in Figure \ref{fig:ae} (b), the contrastive span prediction task aims to force the encoder to learn the representation of the given text by pulling it towards that of its own multiple random spans in the semantic space, while pushing it far away from the representations of all the instances in other groups.
We pre-train the Transformer-based encoder with the contrastive span prediction task and the Masked Language Model (MLM) task.
The pre-trained model, namely COSTA for short, could then be fine-tuned on a variety of downstream dense retrieval tasks.


COSTA enables us to mitigate the aforementioned technical issues since:
(1) COSTA offers incentives for representations of instances in a group sharing the same semantics to be similar, while penalizing the representations of groups expressing different semantics to be distinguished from each other. 
The group-wise supervision enables the encoder to look beyond the local structures of input texts and become more aware of the semantics of other groups. 
This contributes to learning discriminative text representations. 
(2) Without an explicit decoder, our method is able to enforce an information bottleneck on the text representations. 
That is, pre-training the encoder alone can avoid the bypass effect of the decoder thoroughly.

We pre-train COSTA based on the English Wikipedia which contains millions of well-formed wiki-articles. 
We then fine-tune COSTA on four representative downstream dense retrieval datasets, including MS MARCO passage ranking task, MS MARCO document ranking task, and two TREC 2019 Deep Learning tracks. 
The empirical experimental results show that COSTA can achieve significant improvements over the state-of-the-art baselines. 
We also visualize the query and document representations to illustrate how COSTA generates discriminative text presentations to improve the retrieval performance. 
Under the low-resource setting, we show that good retrieval performance can be achieved across different datasets by fine-tuning the COSTA with very little supervision.

\section{Related Work}

In this section, we briefly review the most related topics to our work, including dense retrieval and the pre-training method for IR.

\subsection{Dense Retrieval}

\begin{figure*}[t]
    \centering
    \includegraphics[scale=0.35]{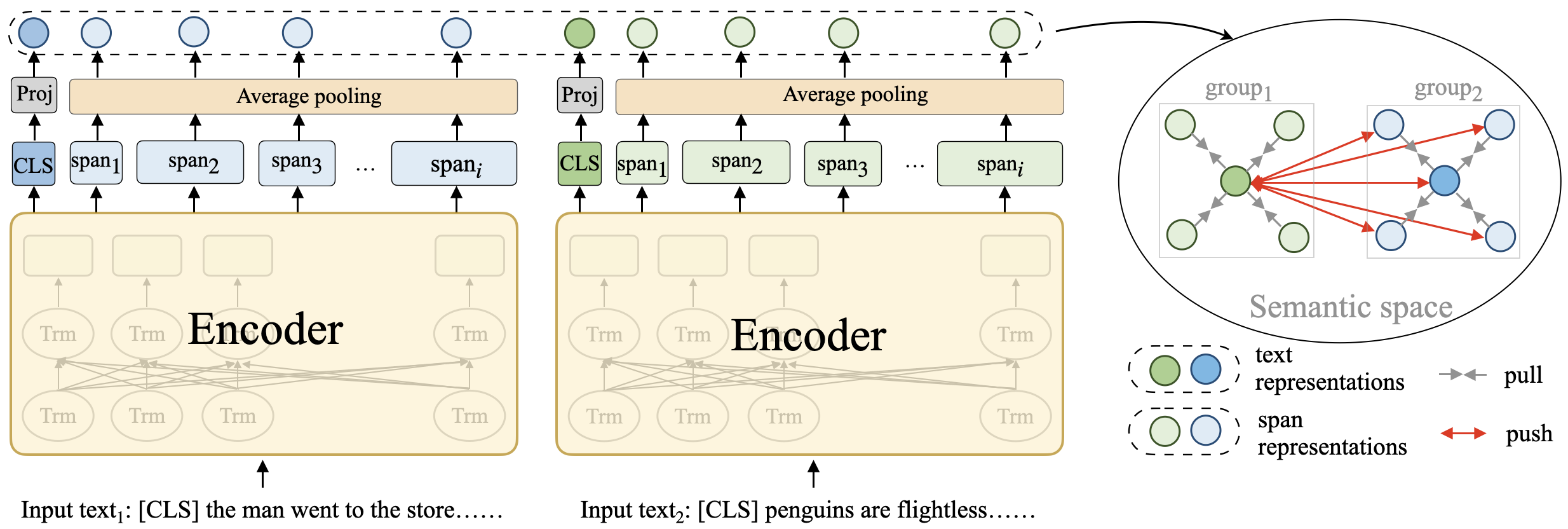}
    \caption{Contrastive Span Prediction Task. A special token [CLS] is added in front of each input text to represent the whole text. The encoder maps the whole text and its own multi-granularity spans into semantic representations. Note a text and its spans form a group. Then, a projector network and an average pooling layer are applied on top of the text representation and the span representation, respectively. Finally, we train the Transformer-based encoder via a group-wise contrastive loss to force the representation of the text close to that of its spans in a group, while far away from other groups.}
    \label{fig:model}
\end{figure*}

Dense retrieval models generally adopt a bi-encoder architecture to encode queries and documents separately for effective search.
The relevance is computed by the simple similarity function like cosine or dot-product.
\citet{Karpukhin2020DensePR} showed that by utilizing in-batch negatives, dense retrieval models perform much better than BM25.
After that, researchers began to explore various fine-tuning techniques to enhance dense retrieval models, such as mining hard negatives~\cite{Xiong2021ApproximateNN,Zhan2021OptimizingDR}, late interaction~\cite{Khattab2020ColBERTEA}, distill knowledge from a strong teacher~\cite{Lin2020DistillingDR}, query clustering~\cite{Hofsttter2021EfficientlyTA}, and data augmentation~\cite{Qu2021RocketQAAO}.
For example, \citet{Xiong2021ApproximateNN} proposed to construct hard negatives from an Approximate Nearest Neighbor (ANN) index of the corpus and period refresh the index during training.
\citet{Zhan2021OptimizingDR} presented a dynamic hard negatives mining method applied upon fine-tuned dense retrieval models.
\citet{Khattab2020ColBERTEA} introduced a MaxSim late interaction operation to model the fine-grained similarity between queries and documents.
\citet{Lin2020DistillingDR} proposed to distill from ColBERT's MaxSim operator into a simple dot product to enable a single-step ANN search.
\citet{Qu2021RocketQAAO} introduced three training strategies, i.e.,  cross-batch negatives, denoised hard negatives, and data augmentation.
Although these methods greatly improve the performance of dense retrieval models, they usually need more computational or storage cost that may limit their use.

\subsection{Pre-training for IR}

Researchers in the IR community have also designed pre-training objectives tailored for IR. 
For example, \citet{Chang2020PretrainingTF} proposed three tasks that closely resemble passage retrieval in question answering (QA):
(1) ICT: The query is a sentence randomly drawn from the passage and the document is the rest of the sentences; 
(2) BFS: The query is a random sentence in the first section of a Wikipedia page, and the document is a random passage from the same page; 
and (3) WLP: The query is a random sentence in the first section of a Wikipedia page, and the document is a passage from another page where there is a hyperlink between the two pages.
When applying these tasks in ad-hoc retrieval, we observed marginal improvements on typical benchmark datasets.
\citet{Ma2021PROPPW, Ma2021BPROPBP} proposed to sample representation words from the document according to a unigram document language model or a contrastive term distribution, and then pre-train the Transformer model to predict the pairwise preference between the two sampled word sets jointly with MLM.
This task is designed for the re-ranking stage, when applied in the retrieval stage, improvements are limited.
\cite{Gao2021CondenserAP,Gao2021UnsupervisedCA} modify the Transformer architecture to establish structural readiness by doing LM pre-training actively condition on dense representations.
The most related work with ours is SEED-Encoder~\cite{lu-etal-2021-seed} that pre-trains an autoencoder with a weak decoder to learn document representations for dense retrieval.
Concretely, they use a three-layer Transformer as the decoder and also restrict its span attention to 2.
But this method has two major issues as we discuss in Intro~\ref{sec:intro}.

\section{Our Approach}

In this section, we describe our proposed pre-training method for dense retrieval in detail.

\subsection{Motivation}

Existing work has demonstrated that autoencoder-based language models are able to learn high-quality text representations for effective search. 
These models typically pair a decoder on top of the encoder and train the decoder to reconstruct the input texts solely from the encoder's encodings. 
However, as observed in previous works \cite{lu-etal-2021-seed}, the decoder in autoencoder-based language models can exploit natural language patterns via its access to previous tokens and bypass the dependency on the encoder, especially when the sequence is long and the decoder is strong. 
Although \citet{lu-etal-2021-seed} has proposed to pre-train the autoencoder using a weak decoder, the bypass effect of the decoder still remains. 
Besides, decoding all the input texts equally may decrease the discriminative power of the representations, since the decoder favors the generation of common words with high frequency.

To address these issues, we propose to discard the decoder and enforce the information bottleneck on the encoder itself for better representation quality. 
We introduce a novel contrastive span prediction task to pre-train the encoder alone. 
Specifically, given an input text, we sample a set of spans to build a group. 
Then, we encourage the representation of the text and that of its own spans in the semantic space to be closer while keeping other groups away. 
In this way, it not only eliminates the bypass effect of the decoder thoroughly, but also aids the pre-trained model in better capturing discriminative semantics of the text via more effective group-level contrastive supervision. 
COSTA is pre-trained with contrastive span prediction task and Masked Language Model (MLM) task. 
The overall model architecture is shown in Figure~\ref{fig:model}. 
The pseudo pre-training algorithm is described in Algorithm~\ref{algo}. 
We introduce the contrastive span prediction task in detail next.

\subsection{Contrastive Span Prediction Task}

Given a mini-batch of input texts, we first sample a set of multi-granularity spans for each text and encode the whole texts and spans into semantic representations. 
A group-wise contrastive loss is then applied to force the whole text representation close to its own random spans while far away from other groups in the same mini-batch.
The detailed pre-training procedures are as follows.

\textbf{Multi-granularity Span Sampling.} Different granularities of span capture different properties of the input text. 
For example, fine-grained spans can capture specific words or entities mentioned in the text.
Coarser-grained spans can instead capture more abstract properties of the text. 
In this work, to better capture the semantic information, we explicitly sample a set of spans at several levels of granularity for each input text, including word-level, phrase-level, sentence-level and paragraph-level. Given a tokenized document with $n$ words $d={(x_1, x_2,...,x_n)}$, the detailed span sampling process is shown as follows.

\begin{enumerate}[leftmargin=*]
    \item For each level of granularity, we first sample the span length from a beta distribution following \cite{Giorgi2021}, i.e., 
\begin{equation}\label{algo:beta}
\begin{aligned}
& p_{span} \sim Beta(\alpha, \beta), \\
& \ell_{span} = p_{span}*(\ell_{max}-\ell_{min}) + \ell_{min},
\end{aligned}
\end{equation}
where $\ell_{min}$ and $\ell_{max}$ denote the minimum length and maximum span length of each level of granularity. $\alpha$ and $\beta$ are two hyperparameters.

\item We then randomly sample the starting position ${start}$ with respect to 
\begin{equation}\label{algo:start}
    {start} \sim U(1,n-\ell_{span}).
\end{equation}

The span is finally determined by the start position ${start}$ and span length $\ell_{span}$:
\begin{equation}\label{algo:span}
\begin{aligned}
& end = start+\ell_{span}, \\
& span = [x_{start}, \dots, x_{end-1}].
\end{aligned}
\end{equation}
We only store the starting and end positions for each span, not the raw span text.
Note that for the word-level span, we sample a whole word (without tokenized) and filter out the stopwords. 
\end{enumerate}

We sample $T$ spans for each level of granularity and repeat the above process for all the granularity to obtain the final $4T$ spans with respect to each input text.

\textbf{Text Encoding.}  To obtain the representations of the whole text and its multiple spans, we adopt the multi-layer Transformer encoder architecture as the text encoder.

Specifically, a special token $x_0 = \mathsf{[CLS]}$ is added in front of the input to represent the whole text like in BERT: ($d={x_0, x_1, x_2,...,x_n}$).
The Transformer encoder maps each word in the $d$ to low-dimensional dense representations, i.e.,
\begin{equation}\label{algo:rep}
\textbf{h}_0, \textbf{h}_1,...,\textbf{h}_n = Transformer(x_0, x_1, ...,x_n), h_i \in \mathbb{R}^H,
\end{equation}
where $H$ is the hidden size.

Previous works \cite{Chen2020ASF,Chen2021ExploringSS,Grill2020BootstrapYO} have suggested that the additional nonlinear projection is critical to prevent representation collapse for contrastive learning.
Therefore, for the whole text representation, we feed it into a projector network $projector(\cdot)$,  which is a feed-forward neural network (FFN) with a non-linear activation, i.e., 
\begin{equation}
    projector(\cdot) = Tanh(FFN(\cdot)).
\end{equation}
Therefore, the whole text representation is transformed into $\textbf{z}_0$ by, 
\begin{equation}\label{algo:drep}
    \textbf{z}_0 = projector(\textbf{h}_0).
\end{equation}

For the spans in the text $d$, we conduct the average pooling operation over the output word representations to obtain the span representations $\{\textbf{z}_i\}_{i=1}^{4T}$.
For example, if the start position and the end position of a span are $start$ and $end$ respectively, the span representation is obtained as follows: 
\begin{equation}\label{algo:span_rep}
    \textbf{z}_i = \text{AvgPooling}([\textbf{h}_{start},...,\textbf{h}_{end-1}]).
\end{equation}

\textbf{Group-wise Contrastive Loss.} Based on the representations of each text and its multi-granularity span representations, we apply a group-wise contrastive loss to pre-train the text encoder.
The group-wise contrastive loss aims to pull the text representation close to the instances in its group, and push it away from the representations from other groups. 
The classical contrastive loss function is incapable of handling such a case where multiple examples instead of one example are denoted as positive to one instance~\cite{Khosla2020SupervisedCL}. 

Specifically, given a mini-batch with $N$ input texts ($d_1,...,d_N$), we can obtain $N$ whole text representations and $N*4T$ span representations, which form a total of $N*(4T+1)$ representations.
Let $S(i)$ be the index set of spans from $d_i$, and the size of $S(i)$ equals $4T$. 
The group-wise contrastive loss function $\mathcal{L}_{GWC}$ is defined as, 
\begin{equation}\label{ctl_loss}
    \mathcal{L}_{GWC} = \sum_{i=1}^{N}- \frac{1}{4T} \sum_{p\in S(i)} log \frac{exp(sim(\textbf{z}_i, \textbf{z}_p)/\tau)}{\sum_{j=1}^{N*(4T+1)}{\mathbbm{1}_{[i\neq{j}]}exp(sim(\textbf{z}_i, \textbf{z}_j)/\tau)}},
\end{equation}
where $sim(\cdot)$ is the dot-product function and $\tau$ is the temperature hyperparameter.

\begin{algorithm}[t]
\caption{Pre-training COSTA}
\label{algo}
\LinesNumbered

Initialize COSTA model parameters $\Theta$. \\
The pre-training corpus $\mathcal{D}$. \\
Set the number of spans sampled per granularity: $T$. \\
Set the max number of epoch: $epoch_{max}$.

\BlankLine

// \textit{multi-granularity span sampling} \\
\For{d in $\mathcal{D}$}{

Sample $4T$ spans for $d$, w.r.t Eq.(\ref{algo:beta}),Eq.(\ref{algo:start}),Eq.(\ref{algo:span}) \\
Pack the span indexes (start, end) into $\mathcal{D}_s$.
}

\BlankLine
// \textit{model pre-training} \\
\For{epoch in 1,2, ..., $epoch_{max}$}{
    \For{mini-batch $b$ in $\mathcal{D}$}{
    // \textit{encode texts into dense representations} \\
    \{\textbf{h}$_0$,\textbf{h}$_1$,...\}$_{0}^{|b|}$ = Eq.(\ref{algo:rep})  \\
    // \textit{obtain the whole text representations} \\
    \{\textbf{z}$_0$\}$_{0}^{|b|}$ = Eq.(\ref{algo:drep}) \\
    // \textit{obtain the span representations} \\
    \{$(start, end)*4T$\}$_{0}^{|b|}$ = $\mathcal{D}_s(b)$ \\
    \{\textbf{z}$_i*4T$\}$_{0}^{|b|}$ = Eq.(\ref{algo:span_rep}) \\
    // Compute loss: L($\Theta$) \\
    L($\Theta$) = Eq.~\ref{ctl_loss} for contrastive span prediction task \\
    L($\Theta$) = Eq.~\ref{mlm} for masked language modeling task \\
    Update model parameters $\Theta$
}   
}
\end{algorithm}

\subsection{Masked Language Modeling}

The MLM \cite{Taylor1953ClozePA} task first randomly masks out some tokens from the input and then trains the model to predict the masked tokens based on the rest of the tokens. 
As discussed in previous works \cite{Ma2021PROPPW}, the MLM objective could in general contribute to building good contextual text representations for IR, thus it would help COSTA to learn good span representations.
Therefore, similar to BERT, we also adopt MLM as one of the pre-training objectives besides the contrastive span prediction objective.

Specifically, the MLM loss $\mathcal{L}_{MLM}$ is defined as: 
\begin{equation}\label{mlm}
\begin{aligned}
\mathcal{L}_{MLM} = -\sum_{\hat{x}\in m(\textbf{x})} \log p(\hat{x}|\textbf{x}_{\backslash m(\textbf{x})}),
\end{aligned}
\end{equation}
where $\textbf{x}$ denotes the input sentences, $m(x)$ and $\textbf{x}_{\backslash m(\textbf{x})}$ denotes the masked words and the rest words from $\textbf{x}$, respectively.

We then pre-train the Transformer-based encoder using the group-wise contrastive loss in Eq. (\ref{ctl_loss}) jointly with the MLM loss in  Eq. (\ref{mlm}), as follows, 
\begin{equation}
\label{final_loss}
\nonumber
\mathcal{L}_{total}  = \lambda\mathcal{L}_{GWC}  + \mathcal{L}_{MLM},  
\end{equation}
where $\lambda$ is the hyperparameter.

\section{Experimental Settings}

We first introduce our experimental settings, including pre-training corpus, downstream tasks, baseline methods, evaluation metrics, and implementation details.

\subsection{Research Questions}
We conduct experiments to verify the effectiveness of COSTA.
Specifically, we target the following research questions: 
\begin{itemize}[leftmargin=*]
\item \textbf{RQ1:} How does COSTA perform compared with baselines using the same fine-tuning strategy? 
\item \textbf{RQ2:} How does COSTA perform compared with advanced dense retrieval models using complicated fine-tuning strategies?
\item \textbf{RQ3:} How do different components of COSTA affect the retrieval performance?
\item \textbf{RQ4:} How about the discriminative ability of COSTA?
\item \textbf{RQ5:} How does COSTA perform under the low-resource setting?
\end{itemize}

\subsection{Pre-training Corpus and Downstream Tasks}
We use the English Wikipedia as our pre-training corpus following the existing work~\cite{Devlin2019BERT,Ma2021BPROPBP,lu-etal-2021-seed,Ma2021PROPPW}.

\begin{itemize}[leftmargin=*]
\item \textbf{Wikipedia} is a widely used corpus that contains millions of well-formed Wiki-articles.

\end{itemize}

We fine-tune COSTA on several standard dense retrieval benchmarks.

\begin{itemize}[leftmargin=*]
\item \textbf{MS MARCO Passage Ranking (MARCO Dev Passage)}~\cite{Campos2016MSMA} is a large-scale benchmark dataset for web passage retrieval, with about 0.5 million training queries and 8.8 million passages.
\item \textbf{MS MARCO Document Ranking (MARCO Dev Doc)}~\cite{Campos2016MSMA} is another large-scale benchmark dataset for web document retrieval, with about 0.4 million training queries and 3 million documents.
\item \textbf{TREC 2019 Passage Ranking (TREC2019 Passage)}~\cite{Craswell2020OverviewOT} replaces the test set in the MS MARCO passage ranking task with a novel set produced by TREC with more comprehensive labeling. 
\item \textbf{TREC 2019 Document Ranking (TREC2019 Document)}~\cite{Craswell2020OverviewOT} replaces the test set in the MS MARCO document ranking task with a novel set produced by TREC with more comprehensive labeling. 
\end{itemize}

\subsection{Baselines}
We adopt three types of baselines for comparison, including sparse retrieve models, other pre-trained models, and advanced dense retrieval models that fine-tune the existing pre-trained models using complicated techniques.

\subsubsection{Traditional Sparse Retrieval Models}
For traditional retrieval models, we consider taking BM25 and DeepCT as the baselines, and we also report several representative results according to the TREC overview paper~\cite{Craswell2020OverviewOT}.

\begin{itemize} [leftmargin=*]
\item \textbf{BM25} \cite{Robertson2009ThePR} is a highly effective strong baseline model that represents the traditional sparse retrieval models. 

\item \textbf{DeepCT}~\cite{Dai2019ContextAwareST,Dai2020ContextAwareDT} is a neural term weighting framework that learns to map BERT's contextualized text representations to context-aware term weights for sentences and passages.

\end{itemize}

\subsubsection{Other Pre-training Models}
Other pre-training methods include the general-purpose pre-trained language model BERT and other pre-trained models tailored for IR:

\begin{itemize} [leftmargin=*]
\item \textbf{BERT} \cite{Devlin2019BERT} is the dominant pre-trained model which achieves great success on various language understanding tasks.
BERT is pre-trained with MLM and Next Sentence Prediction(NSP) tasks using the Transformer encoder.


\item \textbf{ICT}~\cite{Chang2020PretrainingTF} is specifically designed for passage retrieval in QA scenario.
ICT randomly samples a sentence from a passage in a Wiki page and takes the rest sentences in the passage as its positive context.

\item \textbf{PROP}~\cite{Ma2021PROPPW} is designed for ad-hoc retrieval which pre-trains the Transformer model with representative words prediction task(ROP) and MLM.
ROP predicts the pairwise preference for the sampled pairs of word sets from the unigram language model.

\item  \textbf{B-PROP}~\cite{Ma2021BPROPBP} improves PROP by replacing the unigram language model with a contrastive term distribution obtained from the [CLS]-token attention of BERT.

\item \textbf{SEED}~\cite{lu-etal-2021-seed} pre-trains the autoencoder using a weak decoder for dense retrieval.
SEED is pre-trained with the reconstruction task and MLM on English Wikipedia.


\end{itemize}

\subsubsection{Advanced Dense Retrieval Models that Fine-tunes with Complicated Techniques}

As many works have studied various complicated fine-tuning methods to enhance dense retrieval models, we thus also compare COSTA fine-tuned using simple strategies with those models.
These models include:

\begin{itemize} [leftmargin=*]
\item  \textbf{ANCE}\cite{Xiong2021ApproximateNN} investigated the hard negative mining problem where they periodically refresh its corpus index to retrieve the hard negatives produced by the current model.

\item \textbf{TCT-ColBERT}\cite{Lin2020DistillingDR} distills from ColBERT’s MaxSim operator into a simple dot-product operation for computing relevance scores.

\item \textbf{TAS-B}\cite{Hofsttter2021EfficientlyTA} proposes to train dense retrieval models with a query topic-aware and balanced margin of passage pairs sampling strategy.

\item \textbf{ADORE+STAR}\cite{Zhan2021OptimizingDR} is built on the STAR model which has mined one-round hard negatives besides BM25 negatives to train a strong document encoder.
ADORE then retrieves the hard negatives from the pre-build STAR document index in real-time and only optimizes the query encoder.


\item  \textbf{RocketQA}\cite{Qu2021RocketQAAO} utilizes three training strategies, namely cross-batch negatives, denoised hard negatives, and data augmentation, to improve the dense retrieval model.

\end{itemize}

\subsection{Evaluation Metrics}

We use the official metrics of these four benchmarks.
For the MS MARCO passage ranking task, we report the Mean Reciprocal Rank at 10 (MRR@10) and recall at 1000 (R@1000).
For the MS Document ranking task, we report the MRR@100 and R@100.
For two TREC tasks, we report normalized discounted cumulative gain at 10 (NDCG@10), and R@100 and R@1000 for passage ranking and document ranking, respectively.

\subsection{Implementation Details}\label{sec:finetune}

Here, we describe the implementation of pre-training procedures, and fine-tuning procedures in detail.

\subsubsection{Pre-training Procedures}
In this section, we describe the muti-granularity span sampling and model pre-training in detail. 

\noindent%
\textbf{Muti-granularity Span Sampling.}
In the span sampling phase, for phrase-level span, $\ell_{min}$ and $\ell_{max}$ are set to 4 and 16, respectively.
For sentence-level span, $\ell_{min}$ and $\ell_{max}$ are set to 16 and 64, respectively.
For paragraph-level span, $\ell_{min}$ and $\ell_{max}$ are set to 64 and 128, respectively.
The $\alpha$ and $\beta$ hyperparameter in beta distribution is set to 4 and 2, respectively, which skews sampling towards longer spans.
We sample 5 spans for each granularity per input text.

\noindent%
\textbf{Model Pre-training.} Our text encoder uses the same model architecture as BERT~\cite{Devlin2019BERT}.
Considering the large computational cost of pre-training from scratch, we initialize the encoder from the official BERT’s checkpoint and only learn the projector from scratch.
We use a learning rate of 5e-5 and Adam optimizer with a linear warm-up technique over the first 10\% steps.
We pre-train on Wikipedia for 6 epochs.
The long input documents are truncated to several chunks with a maximum length of 512.
The hyper-parameter of $\tau$ in Eq.~\ref{ctl_loss} is set to 0.1.
The hyper-parameter of $\lambda$ in Eq.~\ref{final_loss} is set to 0.1.
We pre-train COSTA on up to four NVIDIA Tesla V100 32GB GPUs.

For BERT and SEED-Encoder, we directly use the open-sourced models as both of them are pre-trained on English Wikipedia and BookCorpus~\cite{Zhu2015AligningBA}.
For ICT, we use the same pre-training procedures as COSTA and pre-train it with MLM and ICT tasks. 
We pre-train PROP and B-PROP using a bi-encoder architecture for a fair comparison.

\subsubsection{Fine-tuning Procedures}
For the passage ranking datasets and the document ranking datasets, we use a similar but not identical fine-tuning strategy.
Following STAR~\cite{Zhan2021OptimizingDR}, we use a two-stage strategy to fine-tune downstream dense retrieval tasks with Tevatron toolkit~\cite{Gao2022TevatronAE}.

\noindent%
\textbf{Fine-tuning Passage Ranking Datasets.}
For the two passage ranking datasets including MARCO Dev Passage and TREC2019 Passage, we train COSTA with official BM25 negatives first for 3 epochs, and then mine hard negatives of the BM25 warm-up model to continue training 2-3 epochs.
Note that we only mine the hard negatives from the BM25 warm-up model once.
The query length and the passage length are set to 32 and 128 respectively.
We use a learning rate of 5e-6 and a batch size of 64.

\noindent%
\textbf{Fine-tuning Document Ranking Datasets.}
For the two document ranking datasets on MARCO Dev Doc and TREC2019 Doc, following existing works~\cite{Xiong2021ApproximateNN, Zhan2021OptimizingDR,lu-etal-2021-seed}, we use the model fine-tuned on the passage ranking task as the starting point.
Since the fine-tuned COSTA is too strong, continuing fine-tuning this model with official BM25 negatives decreases the performance greatly on the document ranking datasets.
We thus iteratively mine the static hard negatives twice and only fine-tune 1 epoch on the static hard negatives for each iteration.
We use a learning rate of 5e-6 and a batch size of 64.
The document length is truncated to the first 512 tokens.

We pair each positive example with 7 negative examples for all these 4 datasets.
All fine-tuning procedures use Adam optimizer with a linear warm-up technique over the first 10\% steps.

\begin{table}[t]
\renewcommand{\arraystretch}{1.2}
  \setlength\tabcolsep{5pt} 
  \caption{Comparisons between COSTA and the baselines on the two passage ranking datasets. Two-tailed t-tests demonstrate the improvements of COSTA to the baselines are statistically significant ($p \le 0.05$). $\ast$ indicate significant improvements over BERT. $\dag$ indicate significant improvements over ICT, PROP, and B-PROP. And $\ddag$ indicate significant improvements over SEED. Results not available or not applicable are marked as `-'.}
  \label{tab:main_res}
  \small
  \begin{tabular}{lllll}
  \toprule

     \multirow{2}{*}{Model} & \multicolumn{2}{c}{MARCO Dev Passage} & \multicolumn{2}{c}{TREC2019 Passage} \\ 
    \cmidrule(lr){2-3} \cmidrule(lr){4-5}
     &  MRR@10 & R@1000 & NDCG@10 & R@1000 \\ 
    \midrule
    \multicolumn{5}{c}{\textit{Sparse retrieval models}} \\
    \midrule
    BM25 & 0.187 & 0.857 & 0.501 & 0.745 \\
    DeepCT\cite{Dai2019ContextAwareST} & 0.243 & 0.905 & 0.551 & - \\
    Best TREC Trad\cite{Craswell2020OverviewOT} & - & - & 0.554 & - \\
    \midrule
    \multicolumn{5}{c}{\textit{Fine-tuning with official BM25 negatives}} \\
    \midrule
    BERT & 0.316 & 0.941 & 0.616 & 0.704   \\
    ICT & 0.324 & 0.938 & 0.618 & 0.705 \\
    PROP & 0.320 & 0.948 & 0.586 & 0.709 \\
    B-PROP & 0.321 & 0.945 & 0.603 & 0.705 \\
    SEED\cite{lu-etal-2021-seed} & 0.329 & 0.953 & - & - \\
    SEED(ours) & 0.331$^{\ast}$ & 0.950$^{\ast}$ & 0.625$^{\ast}$ & 0.733$^{\ast\dag}$ \\
    COSTA & \textbf{0.342}$^{\ast\dag\ddag}$ & \textbf{0.959}$^{\ast\dag}$ & \textbf{0.635}$^{\ast\dag\ddag}$ & \textbf{0.773}$^{\ast\dag\ddag}$ \\
    \midrule
    \multicolumn{5}{c}{\textit{Fine-tuning with static hard negatives}} \\  
    \midrule
    BERT & 0.335 &  0.957 & 0.661 & 0.769  \\
    ICT & 0.339 & 0.955 & 0.670 & 0.775\\
    PROP & 0.337 & 0.951 & 0.673 & 0.771  \\
    B-PROP & 0.339 & 0.952 & 0.672 & 0.774 \\
    SEED & 0.342$^{\ast}$ & 0.963 & 0.679$^{\ast}$ & 0.782$^{\ast\dag}$ \\
    COSTA & \textbf{0.366}$^{\ast\dag\ddag}$ & \textbf{0.971}$^{\ast\dag}$ & \textbf{0.704}$^{\ast\dag\ddag}$ & \textbf{0.816}$^{\ast\dag\ddag}$ \\
    \bottomrule
  \end{tabular}
\end{table}
\begin{table}[t]
\renewcommand{\arraystretch}{1.3}
\setlength\tabcolsep{3pt} 
  \caption{Comparisons between COSTA and the baselines on two document ranking datasets. Two-tailed t-tests demonstrate the improvements of COSTA to the baselines are statistically significant ($p \le 0.05$). $\ast$ indicates significant improvements over BERT. $\dag$ indicates  significant improvements over ICT, PROP, and B-PROP. $\ddag$ indicates  significant improvements over SEED. Results not available or not applicable are marked as `-'.}
  \label{tab:main_res2} 
  \begin{tabular}{lllll}
  \toprule

     \multirow{2}{*}{Model} & \multicolumn{2}{c}{MARCO Dev Doc} & \multicolumn{2}{c}{TREC2019 Doc}  \\ 
    \cmidrule(lr){2-3} \cmidrule(lr){4-5} 
     &  MRR@100 & R@100  & NDCG@10 & R@100  \\ 
    \midrule
    \multicolumn{5}{c}{\textit{Sparse retrieval models}} \\
    \midrule
    BM25 & 0.277 & 0.808 & 0.519 & 0.395 \\
    DeepCT\cite{Dai2019ContextAwareST} & 0.320 & - & 0.544 & - \\
    Best TREC Trad\cite{Craswell2020OverviewOT} & - & - & 0.549 & -\\
    \midrule
    \multicolumn{5}{c}{\textit{1st iteration: Fine-tuning with static hard negatives}} \\
    \midrule
    BERT & 0.358 & 0.869 & 0.563 & 0.266 \\
    ICT & 0.364 & 0.873 & 0.566 & 0.273\\
    PROP & 0.361 & 0.871 & 0.565 & 0.269 \\
    B-PROP & 0.365 & 0.871 & 0.567& 0.268\\
    SEED & 0.372$^{\ast}$ & 0.879$^{\ast}$ & 0.573$^{\ast}$ & 0.272  \\
    COSTA  & \textbf{0.395}$^{\ast\dag\ddag}$ & \textbf{0.894}$^{\ast\dag\ddag}$ & \textbf{0.582}$^{\ast\dag\ddag}$ & \textbf{0.278}$^{\ast}$ \\
    
    \midrule
    \multicolumn{5}{c}{\textit{2nd iteration: Fine-tuning with static hard negatives}} \\
    \midrule
    BERT & 0.389 & 0.877 & 0.594 & 0.301 \\
    ICT & 0.396 & 0.882 & 0.605 & 0.303\\
    PROP & 0.394 & 0.884 & 0.596 & 0.298 \\
    B-PROP & 0.395 & 0.883 & 0.601& 0.305\\
    SEED & 0.396 & 0.902$^{\ast}$ & 0.605$^{\ast}$ & 0.307  \\
    COSTA  & \textbf{0.422}$^{\ast\dag\ddag}$ & \textbf{0.919}$^{\ast\dag\ddag}$ & \textbf{0.626}$^{\ast\dag\ddag}$ & \textbf{0.320}$^{\ast\dag\ddag}$ \\
    \bottomrule
  \end{tabular}
\end{table}

\section{Experiment results}

In this section, we analyze the experimental results to demonstrate the effectiveness of the proposed COSTA method. 

\subsection{Baseline Comparison with the Same Fine-tuning Strategy}\label{sec:baseline}
To answer \textbf{RQ1}, we compare COSTA with various baselines on four benchmark datasets. 
For a fair comparison, different pre-trained models leverage the same fine-tuning strategy. 
The performance comparisons are shown in Table \ref{tab:main_res} and Table \ref{tab:main_res2}.

\textbf{Results on the two passage ranking datasets.} As shown in Table \ref{tab:main_res}, we have the following observations: 
(1) The relative order of different models fine-tuned with BM25 negatives is quite consistent with that fine-tuned with static hard negatives.
(2) The dense retrieval models outperform all the traditional sparse retrieval models by a large margin. 
It indicates that by capturing the semantics meanings of queries and documents, dense retrieval models can overcome the vocabulary mismatch problem caused by sparse retrieval models for better ranking performance. 
(3) PROP and B-PROP show slight improvements over BERT, indicating that the  ROP task is not as effective in the retrieval stage as that in the re-ranking stage. 
The reason might be that the ROP task only considers the inside information of a document (i.e., it samples words from the corresponding document language model), leading to the unawareness of the outside information (i.e., other documents).
(4) Although ICT utilizes the in-batch negatives, taking a random sentence as the pseudo query can only learn topic-level relevance which is insufficient for dense retrieval tasks.
(5) SEED performs better than BERT, PROP, B-PROP and ICT significantly indicating that by enforcing a bottleneck on the data, SEED can provide better semantic representations for dense retrieval.

When we look at COSTA, we find that COSTA achieves significant improvements over all the baselines.
(1) COSTA can outperform SEED significantly, demonstrating dropping the decoder and utilizing the contrastive strategy help it to produce better text representations.
(2) Compared with ICT, PROP and B-PROP, COSTA is a unified task since the group-wise contrastive span prediction task teaches the model not only to learn better representations from the input itself but also to be distinguishable from other representations.

\textbf{Results on the two document ranking datasets.} As shown in Table \ref{tab:main_res2}, the performance trend on the document ranking datasets is consistent with the passage ranking datasets except for the TREC2019 Doc dataset.
For the TREC2019 Doc, dense retrieval models significantly underperform BM25 in terms of R@1000.
This phenomenon is also observed in other works~\cite{Xiong2021ApproximateNN,Zhan2021OptimizingDR,lu-etal-2021-seed,Craswell2020OverviewOT}.
A possible reason is that MS MARCO has many unlabeled relevant documents in the corpus, but the model often treats these documents as negatives.
So this problem would bias the process of model training and thus hurt its performance.
Finally, we can see that COSTA achieves more improvement in terms of ranking metrics like MRR and NDCG than the recall metric.
For example, COSTA improves SEED's top-ranking performance by 7\% on both the MS MARCO passage ranking dataset and document ranking dataset in terms of MRR metric.
It shows that COSTA has a better discriminative ability that can rank the positive document higher.

\subsection{Baseline Comparison with Different Fine-tuning Strategies}
To answer \textbf{RQ2}, we further compare COSTA with advanced dense retrieval models that fine-tune existing pre-trained models with complicated strategies.
As shown in Table~\ref{tab:complex_finetuning}, we can observe that surprisingly, fine-tuning with simple strategies COSTA performs better than these advanced dense retrieval models with complicated fine-tuning strategies.
Some dense retrieval models are also based on more powerful pre-trained models like RoBERTa~\cite{Robertson2009ThePR} and ERNIE~\cite{sun2019ernie}.
The better results indicates that the performance improvement of COSTA  mainly comes from our pre-training stage.
This demonstrates the effectiveness of our proposed contrastive span prediction task.
Undoubtedly, these fine-tuning methods can also be applied to our COSTA, and we leave this for further study.

\begin{table}[t]
\renewcommand{\arraystretch}{1.3}
\setlength\tabcolsep{8pt}
  \caption{Comparison between COSTA and advanced dense retrieval models using  complicated fine-tuning strategies on the MARCO Dev Passage. Best results are marked bold.}
  \label{tab:complex_finetuning}
  \begin{tabular}{lccc}
    \toprule
   Model &  MRR@10 & R@1000  \\
    \midrule
    ANCE\cite{Zhan2021OptimizingDR} & 0.330 & 0.959  \\
    TCT-ColBERT\cite{Lin2020DistillingDR} & 0.335 & 0.964 \\
    TAS-B\cite{Hofsttter2021EfficientlyTA} & 0.343 & \textbf{0.976} \\
    ADORE+STAR\cite{Zhan2021OptimizingDR} & 0.347 & - \\
    RoctetQA w/o Data Aug \cite{Qu2021RocketQAAO} &  0.364 & - \\
    COSTA &  \textbf{0.366} & 0.971 \\
  \bottomrule
\end{tabular}
\end{table}

\subsection{Breakdown Analysis}
To answer \textbf{RQ3}, in this section, we ablate several components of the span sampling procedure, the projector network in COSTA, and the temperature hyperparameter in the loss function.
All reported results in this section are based on the MS MARCO passage ranking development set trained with only BM25 negatives.
The default experiment setting is that we sample 3 spans for each level of granularity, use a projector to transform the whole text representations and the temperature is set to 1.
We pre-train the model using the default settings on Wikipedia for 3 epochs (half of our full pre-training schedule).

\subsubsection{Impact of the Span Granularity}
We first study the impact of span granularity in the span sampling procedure.
In order to control the total number of spans sampled per input text unchanged, we set it to 12.
That is, when we sample 3 of 4 span granularity, 4 spans are sampled for each of the other three granularity.

Table~\ref{tab:span_type} shows the impact of the different span granularity.
We can see that all the granularity contributes to the whole text representation learning.
But with the increase of span length, i.e., from word-level to paragraph-level, the performance decreases more and more.
For example, without the word-level span, it leads to a slight decrease in terms of MRR@10 while without the paragraph-level span, it leads to a significant decrease.
This may be because longer spans contain more information of the input and thus contribute more to the text representation learning.
Shorter spans like phrase-level and sentence-level are also helpful as they are useful for capturing fine-grained information of the input texts.

\begin{table}[t]
    \caption{The performance of COSTA with different span granularities. Best results are marked bold.}
\renewcommand{\arraystretch}{1.3}
    \setlength\tabcolsep{8pt}
    \centering
    \begin{tabular}{lcc}
    \toprule
        Method & MRR@10 & R@1000 \\
    \midrule
        Base & \textbf{0.335} & \textbf{0.952} \\
        w/o word-level & 0.334 & 0.952 \\
        w/o phrase-level & 0.331  & 0.953\\
        w/o sentence-level & 0.331 & 0.947 \\
        w/o paragraph-level & 0.326 & 0.940 \\
        \bottomrule
    \end{tabular}

    \label{tab:span_type}
\end{table}

\begin{table}[h]
\renewcommand{\arraystretch}{1.3}
    \setlength\tabcolsep{6pt}
 \caption{Performance comparison of COSTA with different span numbers. Best results are marked bold.}
    \centering
    \begin{tabular}{lcccc}
    \toprule
        Span Number & 3 & 5 & 10 & 20 \\
    \midrule
       MRR@10  & 0.335 & \textbf{0.339} & 0.332 & 0.320 \\
       R@1000  & 0.952 & \textbf{0.953} & 0.949 & 0.946 \\
        \bottomrule
    \end{tabular}
    \label{tab:span_num}
\end{table}


\begin{table}[t]
    \caption{Performance comparison of COSTA with different projector architectures in the pre-training and fine-tuning phase. Best results are marked bold.}
    \renewcommand{\arraystretch}{1.3}
    \setlength\tabcolsep{5pt}
    \centering
    \begin{tabular}{l|cc|cc}
    \hline
         \multirow{2}{*}{Setting} & \multicolumn{2}{c|}{Pre-training} & \multicolumn{2}{c}{Fine-tuning} \\ 
         \cmidrule(lr){2-3} \cmidrule(lr){4-5} 
         & MRR@10 & R@1000 & MRR@10 & R@1000 \\
    \hline
        w/ nonlinear Proj & 0.335 & 0.952 & 0.335 & 0.951 \\
        w/ linear Proj & 0.332 & 0.950 & 0.334& 0.952 \\
        w/o & 0.327 & 0.944 & 0.335 & 0.952 \\
        \hline
    \end{tabular}

    \label{tab:projector}
\end{table}

\subsubsection{Impact of the Span Number}
We then study the impact of the number of spans sample per input text.
Given a mini-batch containing $N$ input texts, we sample $T$ spans for each granularity which results in $N*4T$ spans.

As shown in Table~\ref{tab:span_num}, we varies $T$ from 3 to 20.
We can see that with more spans sampled from the input texts, the performance doesn't improve consistently.
When $T=5$, it achieves the best performance across other settings.
We hypothesize this is because the difficulty of the group-wise contrastive objective increases especially at the beginning of training.
Recall that the loss function in Eq.(\ref{ctl_loss}) forces the whole text representations close to their random spans, when $T=20$, COSTA needs to align the whole text representations with $4*T=80$ span representations.
So the model may be hard to learn as the text representations are distributed randomly at the beginning.
The difficulty is also observed from the loss curve where the loss decreases very slowly.
Other learning techniques like curriculum learning~\cite{Bengio2009CurriculumL} are left for further studies to alleviate this problem.
The best span number sampled per granularity in our experiment is $T=5$.

\begin{table}[h]
\renewcommand{\arraystretch}{1.3}
    \setlength\tabcolsep{6pt}
 \caption{Performance comparison of COSTA with different temperatures  $\tau$ on MS MARCO Dev Passage. Best results are marked bold.}
    \centering
    \begin{tabular}{lcccc}
    \toprule
        $\tau$ & 10 & 1 & 0.1 & 0.01 \\
    \midrule
       MRR@10  & 0.274 & 0.329 & 0.335 & 0.267 \\
        \bottomrule
    \end{tabular}
   
    \label{tab:temperature}
\end{table}

\subsubsection{Impact of the Projector Architecture}
Existing work in contrastive learning shows that introducing a learnable nonlinear transformation between the representation and the contrastive loss substantially improves the quality of the learned representations~\cite{Chen2020ASF,Chen2021ExploringSS,Grill2020BootstrapYO}.
We thus study three different architectures for the projector in the pre-training phase and the fine-tuning phase respectively: nonlinear MLP, linear MLP, and identity mapping (i.e., without MLP).
When studying the pre-training stage, we discard the projector in the fine-tuning stage.
When studying the fine-tuning stage, we use the model pre-trained with a nonlinear projector network.

The performance comparison is shown in Table~\ref{tab:projector}.
We can observe that in the pre-training phase, using a nonlinear projection perform better than using a linear projection, and much better than no projection.
One possible reason is that introducing a nonlinear projection ensures that the whole text representation is not equal to any span representation since they are output from the same layer, and also makes it easy to align with multiple span representations simultaneously.
We also studied the impact of the projector in the fine-tuning stage.
For the fine-tuning stage, there is no big difference when fine-tuning the model with or without the projector network.
It indicates that the pre-trained model is strong enough to learn high-quality text representations given the task labels for dense retrieval.

\subsubsection{Impact of the Temperature in Loss Function}
The temperature $\tau$ in the contrastive loss function Eq. \ref{ctl_loss} is used to control the smoothness of the distribution normalized by softmax operation and thus influences the gradients when backpropagation.
A large temperature smooths the distribution while a small temperature sharpens the distribution.
As shown in Table\ref{tab:temperature}, we find the performance is sensitive to the temperature.
Either too small or too large temperature will make our model perform badly.
It might be that a smaller value would make the model too hard to converge while a larger one leads to a trivial solution.
We select 0.1 as the temperature in our COSTA pre-training.


\begin{figure}[h]
	\centering
		\includegraphics[scale=0.26]{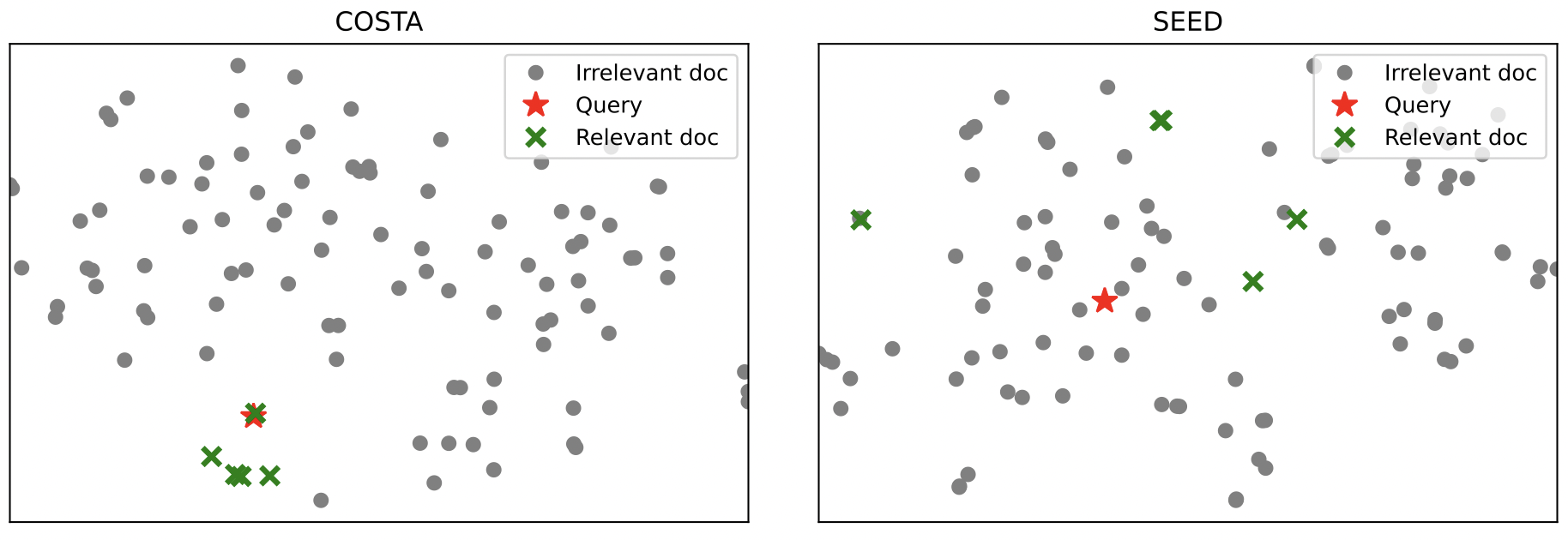}
		  \caption{The t-SNE plot of query and document representations for SEED and COSTA. The QID is 47923 and is from TREC2019 Passage test set.}
  \label{fig:tsne}
\end{figure}

\subsection{Visual Analysis}\label{sec:vis}
To answer \textbf{RQ4}, we visualize the query and text representations using t-SNE to see their distributions in the semantic space.
We conduct this analysis on the MARCO Dev Passage dataset.




In detail, we first fine-tune COSTA and SEED with a very limited number of queries, i.e., 1000, on MS MARCO Dev Passage dataset.
We then plot a t-SNE example using the sampled query and its top-100 candidate documents.
Results in Figure~\ref{fig:tsne} show that COSTA maps the relevant document in the semantic space closer while far away from others.
For SEED, the distribution of relevant documents in the latent space is relatively random.
This is because that SEED only reconstructs the input texts and treats all tokens equally which is unable to learn discriminative representations.
This demonstrates that by pre-training with the contrastive span prediction task, COSTA can generate discriminative dense representations compared with SEED.

\begin{figure}[h]
	\centering
		\includegraphics[scale=0.4]{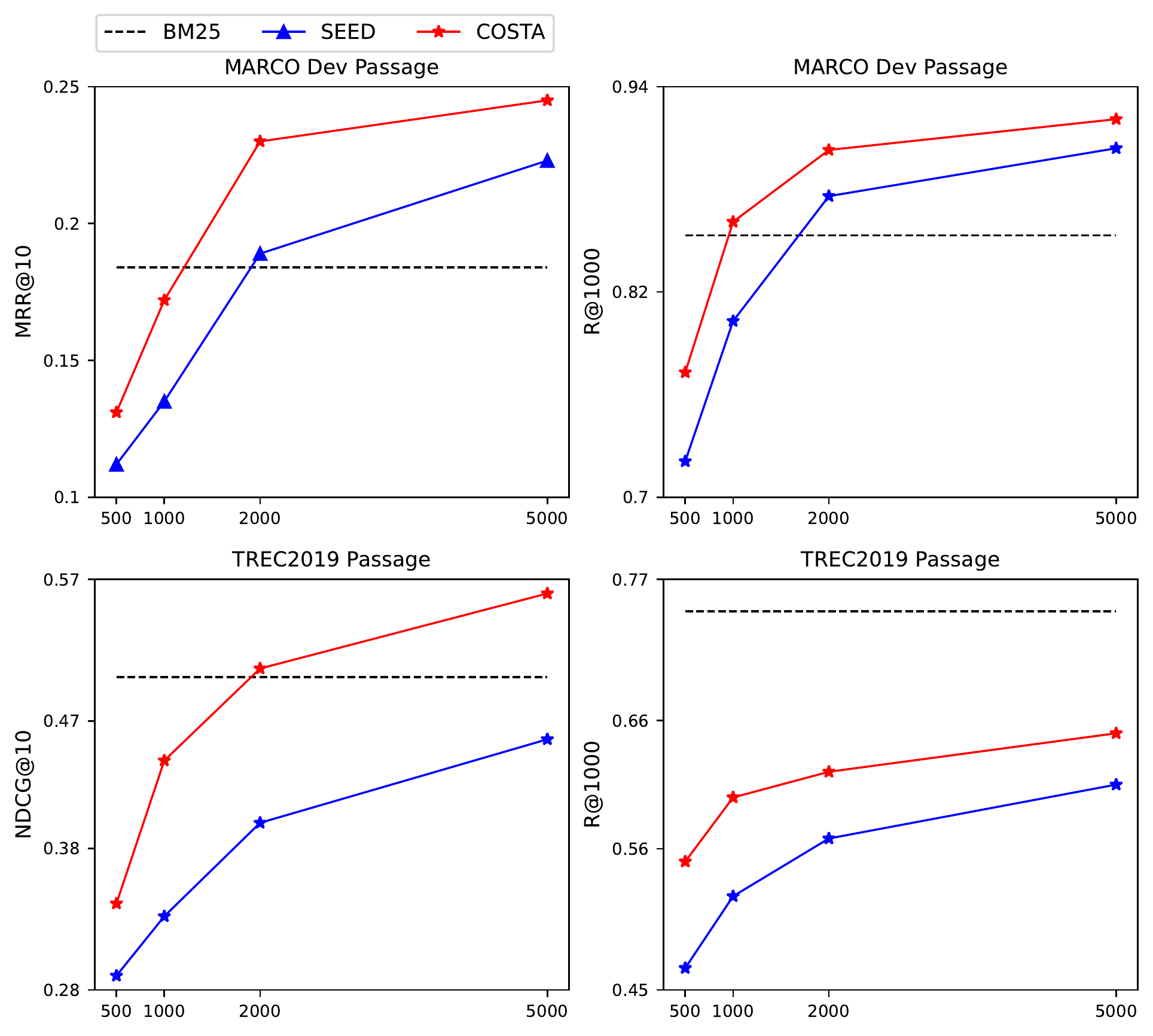}
		  \caption{Fine-tuning with limited supervised data. The x-axis indicates the number of training queries.}
  \label{fig:fs}
\end{figure}

\subsection{Low-Resource Setting}\label{sec:low-resource}
To answer \textbf{RQ5}, we simulated a low-resource setting on the MS MARCO passage ranking dataset and TREC2019 passage ranking dataset.
We randomly sample different fixed limited numbers of queries from the original training set following the existing work~\cite{Gao2021SimCSESC}.
Specifically, we choose to sample 500, 1000, 2000, and 5000 queries for this experiment.
And to improve the stability, we sample 3 different splits for each setting and report the average performance for each setting.
We fine-tune COSTA and SEED with batch size as three different values (i.e., 4, 8, 16), learning rate as two different values (i.e., 5e-6, 1e-5), and pick the last checkpoint to evaluate the original large development set.

As shown in Table~\ref{fig:fs}, we can see that: 
(1) COSTA performs significantly better than SEED on these two datasets in terms of all metrics, indicating that COSTA is able to learn more discriminative text representations than SEED.
(2) Fine-tuning a limited number of training queries, COSTA can outperform the strong BM25 baseline.
For example, COSTA outperforms BM25 with only less than 0.5\% queries, i.e.,2000 queries, on MARCO Dev Passage in terms of both MRR@10 and R@1000.
(3) COSTA performs worse on the TREC2019 Passage dataset since COSTA is fine-tuned on MARCO Dev Passage which suffers the false-negative problem~\cite{Qu2021RocketQAAO}.
\section{Conclusion}

In this paper, we proposed a novel contrastive span prediction task to pre-train a discriminative text encoder for dense retrieval. 
By enhancing the consistency between representations of the original text and its own spans while pushing it away from representations of other groups, COSTA can leverage the merits of both the bottleneck principle and discriminative ability for better representation quality.
The bottleneck principle guarantees the representation can represent itself through ``reconstruction'' and the contrastive strategy can help to make the representation distinguishable across other texts.
Through experiments on 4 benchmark dense retrieval datasets, COSTA outperforms several strong baselines.
Through visualization analysis and the low-resource setting, we demonstrate that COSTA can produce discriminative representations for dense retrieval.
In future work, we would like to apply COSTA to other IR scenarios like open-domain question answering and conversational systems.

\begin{acks}
This work was funded by the National Natural Science Foundation of China (NSFC) under Grants No. 62006218, 61902381, and 61872338, the Youth Innovation Promotion Association CAS under Grants No. 20144310, and 2021100, the Lenovo-CAS Joint Lab Youth Scientist Project, and the Foundation and Frontier Research Key Program of Chongqing Science and Technology Commission (No. cstc2017jcyjBX0059).
\end{acks}


\clearpage
\bibliographystyle{ACM-Reference-Format}
\bibliography{main}

\end{document}